\documentclass[aps,prd,onecolumn,preprintnumbers,floatfix,superscriptaddress,showpacs]{revtex4}

\usepackage{amsmath}
\usepackage{soul}
\usepackage{graphicx}

\newcommand{\tr}{{\rm Tr}}

\newcommand{\zi}{z_{\rm IR}}
\newcommand{\D}{{\cal D}}

\begin{document}

\preprint{ITEP-TH-23/09}

\author{Maxim Konyushikhin}

\affiliation{SUBATECH, Universit\'e de Nantes, 4 rue Alfred Kastler, BP 20722,
Nantes 44307, France}
\affiliation{Institute for Theoretical and Experimental Physics, B.
Cheremushkinskaya 25, Moscow 117259, Russia}

\title{Four-point vector correlators and AdS/QCD correspondence}

\begin{abstract}
We derive the four-point vector correlators in QCD from AdS/QCD correspondence.
It is shown that meson poles are correctly reproduced. The final expression also
suggests a nonzero amplitude in the limit of zero virtuality of two longitudinal
photons. This fact does not mean that one can produce, absorb or
scatter real longitudinal photons.
\end{abstract}

\keywords{QCD, AdS-CFT Correspondence}
\pacs{
11.25.Tq, 
11.10.Kk, 
11.15.Tk  
12.38.Lg  
}

\maketitle

\section{Introduction}

It is known that some strongly coupled gauge theories have a dual
description. In the recent years several effective models of AdS/QCD duality
were suggested. For example, a simple five-dimensional framework in a curved
background introduced in \cite{Erlich:2005qh} appears to be useful for
obtaining low-energy quantities. Namely, in this model with only three free
parameters one can reproduce lightest meson masses, their decay rates and
couplings with a surprisingly good precision. The model is described in the next
section.

In the present paper we derive the four-point vector current correlator in QCD in the
strong coupling limit within AdS/QCD model introduced in \cite{Erlich:2005qh}.

We introduce an additional dilaton field \cite{Karch:2006pv} in order to
reproduce Regge behavior of meson masses at large energies. It is shown that our
result correctly reproduces meson spectrum at small and at large energies. The
expression (\ref{eq5}) suggests the final answer for the QCD four-point vector
correlator.

We also calculate the four-point vector current correlator with
two conserved and two longitudinal currents.
The expression (\ref{eq_par}) shows that the resulting amplitude is finite when
virtuality of longitudinal photons tends to zero. This naively suggests that
there is a nonzero possibility to emit real longitudinal photons. In the
contrary, gauge invariance strongly prohibits the possibility of producing,
absorbing and scattering real longitudinal gauge particles in physical
processes.
Athought the processes involving virtual longitudinal gauge particles exist and
give the essential contribution to the full cross-section, this does not explain
the contradiction.
An analogous problem was already discussed in \cite{Gorsky:1989qd,Smilga:1990uq}
in the context of processes in massless quantum electrodynamics. It was
shown that such a contradiction has a physically sensible explanation. 
On the one hand, it happens that the amplitude in the limit of 
longitudinal photon zero virtuality is not smooth. This implies that one can not
judge about real longitudinal photon interaction by this limit. On the other hand,
although the discussed processes can in principle be observed in deep inelastic
collisions, the smallest experimentally detectable virtuality $p^2$ is of order
$E/L$ ($E$ is photon energy and $L$ is the characteristic apparatus size).

An additional Chern-Simons term can be introduced to the theory
\cite{Hill:2006ei,Grigoryan:2008up,Domokos:2007kt}. In the case of vector
correlator its contribution is zero as it has the form $S_{\rm CS}[A_{\rm
L}]-S_{\rm CS}[A_{\rm R}]$ with $A_{\rm L}=A_{\rm R}=V$.

We also derive some particular kinematic limits and show that the result has a
symmetric form. Namely, if two external momenta are on-shell, the resulting
four-point function is symmetric with respect to the exchange of any other two squared
momenta and the $s$ Mandelstamm variable.

In the next section we briefly discuss the model. In the section III the general
form of four-point vector correlator is given. In the section IV we consider some
particular limits, and the conclusions in the section V complete the paper.

\section{The model}
The simple holographic five-dimensional model of QCD suggested in \cite{Erlich:2005qh} describes dual dynamics of left- and right-handed currents corresponding to the
${\rm SU_L}(N_f)\times {\rm SU_R}(N_f)$ chiral flavor symmetry of QCD.
The 5D action
\begin{equation}\label{eq1}
  S=\int d^4xdz\, e^{-\Phi}\sqrt{g}\,\tr\left[\left|\D
X\right|^2+3\left|X\right|^2-\frac{1}{4g_5^2}\left(F_{\rm L}^2+F_{\rm
R}^2\right)\right]
\end{equation}
($\D_\mu X=\partial_\mu X-iA_{{\rm L}\mu} X+iXA_{{\rm R}\mu}$, $g_5=\frac{12\pi^2}{N_f}$),
written in the anti-de Sitter metric,
\begin{equation}
  ds^2=\frac{1}{z^2}\left(-dz^2+dx^\mu dx_\mu\right),
\end{equation}
is defined for three fields.
These are the scalar $X^a$ in bifundamental representation of ${\rm
SU_L}(N_f)\times {\rm SU_R}(N_f)$ and left and right gauge vector fields
$A_{{\rm L}\mu}^a$, $A_{{\rm R}\mu}^a$ (corresponding to chiral left- and right-handed currents).
In these terms, the four-dimensional QCD theory lies on AdS boundary.
It is more convenient for us to use vector $V_\mu^a$ and axial $A_\mu^a$ gauge fields defined
by $A_{\rm L}=V+A$, $A_{\rm R}=V-A$. They describe vector and axial currents in 4D theory. See
\cite{Erlich:2005qh} for details.

The model at hand must include in some way a parameter of dimension of mass responsible to chiral symmetry breaking. It is its presence that allows one to reproduce low energy physics.
We consider two simplest ways to introduce this parameter.
The hard-wall model \cite{Erlich:2005qh} corresponds to the special case when the dilaton
field $\Phi$ is zero and the ${\rm AdS}_5$ space is cut at some ``infrared'' point
$z=\zi$ ($\zi=1/323$ MeV). The second case corresponds to the so called soft-wall model
when the external dilaton field $\Phi=\Lambda^2 z^2$ ($\Lambda\sim 300$ MeV) gives the scale parameter for the theory. It provides the correct meson Regge trajectories \cite{Karch:2006pv}.
We derive all expressions in the general way
so that they can be applied for both models. Some of their differences are
compared in section IV.

Throughout this paper the gauge $A_{{\rm L}z}=A_{{\rm R}z}=0$ is used and
$\mu$, $\nu$, $\alpha$, $\dots$ stand for 4-dimensional indices.

Finally, it is necessary to impose boundary conditions for the fields at hand.
We put $A_{{\rm L}\mu}=A_{{\rm R}\mu}=0$ at $z=0$ and
appropriate boundary conditions at large $z$. For the hard-wall
model we take $\partial_z F_{\rm L}(z=\zi)=\partial_z F_{\rm R}(z=\zi)=0$; for the
soft-wall model one should take the natural boundary conditions at $z=\infty$ to
make the action finite. See below about the $X$ field boundary conditions.

Let us write the classical quadratic equations of motion for left and right
vector fields ($A_{\rm L}$ or $A_{\rm R}$):
\begin{equation}\label{eq3} 
\left(\partial_z\frac{e^{-\Phi(z)}}{z}\partial_z-\frac{e^{-\Phi(z)}}{z}
\partial^2\right)A_\nu^a+\frac{e^{-\Phi(z)}}{z}\partial_\nu\partial_\mu
A_\mu^a=0.
\end{equation}
 In this way, using the appropriate boundary
conditions, the solution to the equation (\ref{eq3}) correctly reproduces meson
spectrum. (Whereas the hard-wall model reproduces correctly only low-lying meson
spectrum.) Namely, we introduce vector field propagator (in the 4D momentum
representation)
\begin{equation}
  G_{\mu\nu}(k, z,z')=-\left(g_{\mu\nu}-\frac{k_\mu
k_\nu}{k^2}\right)G_{k^2}(z,z')+\frac{k_\mu k_\nu}{k^2}G_0(z,z'),
\end{equation}
\begin{equation}\label{eq4}
  \left[\partial_z \frac{e^{-\Phi(z)}}{z}\partial_z+k^2\frac{
e^{-\Phi(z)}}{z}\right] G_{k^2}(z,z')=i\delta(z-z').
\end{equation}
Here $k_\mu$ is a 4D momentum.
Note that longitudinal part of $G_{\mu\nu}(k,z,z')$ does not depend on $k^2$.
The $k^2$ poles of $G_{k^2}(z,z')$ correspond to meson masses in this
holographic model and are in good agreement with experimental data.

The vacuum in this theory corresponds to $A_{\rm L}=A_{\rm R}=0$.
Let us tell some words about the $X$ field boundary conditions
\cite{Erlich:2005qh}. In the model without a dilaton field the solution on a
vacuum $X$ state is
\begin{equation}\label{eq_xbound}
X=\frac{1}{2}M z+\frac{1}{2}\Sigma z^3,  
\end{equation}
where the matrix $M$ corresponds to the quark mass matrix and $\Sigma$ can be
viewed as a quark condensate, $\Sigma^{\alpha\beta}=\left<\bar q^\alpha
q^\beta\right>$. In the model with a dilaton field the expression
(\ref{eq_xbound}) must be considered as a boundary condition at UV (small $z$).
We choose the simplest possible case when both matrices are proportional to unit
matrix, $M=m_q \boldsymbol 1$, $\Sigma=\sigma\boldsymbol 1$. In this way, the $X$
field itself is proportional to the unit matrix and {\em drops out from our computations}.

\begin{figure}[h]
  \centering
  \includegraphics[width=6in]{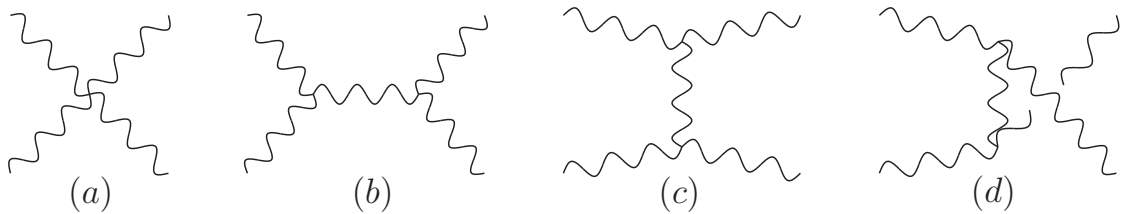}
  \caption{Four-point Feynman graphs as viewed from a 5D theory. Free legs
correspond to ``bulk-to-boundary propagator''.}
  \label{feyn2}
\end{figure}

As the AdS/QCD duality suggests, connected Green's functions in QCD can be
generated by differentiating the classical five-dimensional action (\ref{eq1})
with respect to the sources (UV boundary conditions on $A_{\rm L}(z,x)$ and
$A_{\rm R}(z,x)$).
The classical five-dimensional action can be easily derived by summing the
tree Feynman diagrams.
Particularly, to obtain the four-point vector Green's function we need to sum
over four graphs (see FIG.~\ref{feyn2}).
In order to make the expression finite at the UV boundary (small $z$) we
introduce a cutoff $z=\epsilon\rightarrow 0$ and follow the renormalization
procedure \cite{Bianchi:2001kw}.
Note again that the field $X$ does not interact with the vector field $V_\mu^a$
as their interaction refers to the commutator $\left[V_\mu,X\right]$ with the
vacuum field $X$ which is proportional to the unit matrix. In the derivation of
four-point function of axial vector fields one needs to include an extra diagram
representing an intermediate $X$ field particle exchange.

\section{Vector four-point functions}

\subsection{Conserved currents four-point function}
Let us discuss the first vector four-point function -- the holographic dual of
correlator of four conserved vector currents in 4D theory, i.e. when there are
no sources for the axial field $A^a_\mu(x,z)$ and
the source for the vector field is $V^a_\mu(x,\epsilon)=v_\mu^a(x)$,
$\partial_\mu v_\mu^a(x)=0$. The answer can be easily obtained by summarizing
the Feynman graphs (FIG.~\ref{feyn2}):
\begin{equation}\label{eq_s5}
  S_5(\bot,\bot,\bot,\bot)=\frac{1}{4!}
w_{\alpha\beta\gamma\delta}^{abcd}(\bot,\bot,\bot,\bot)\,\, v_{1\alpha}^a
v_{2\beta}^b v_{3\gamma}^c v_{4\delta}^d,
\end{equation}
where $v_{i\alpha}^a\equiv v^a_\alpha(k_i)$ is a Fourier transformation of
$v_\mu^a(x)$, $k_\mu v_\mu^a(k)=0$. It is not difficult to understand the meaning
of $w_{\alpha\beta\gamma\delta}^{abcd}$. As it is already mentioned,
the differentiation of (\ref{eq_s5}) with respect to the sources $v_\mu^a(k)$
gives the result of calculation of 4D vector four-point function from its
5-dimensional perspective. The explicit expression is
{\small
\begin{multline}\label{eq5}
  w_{\alpha\beta\gamma\delta}^{abcd}(\bot,\bot,\bot,\bot)=\frac{1}{g_5^2}\left\{
  -\left[\int\frac{dz}{z}e^{-\Phi(z)}    V_1(z)V_2(z)V_3(z)V_4(z)\right] \cdot
\right.
\\[2mm]
  \cdot\Big[   
f^{abe}f^{cde}\left(g^{\alpha\gamma}g^{\beta\delta}-g^{\alpha\delta}g^{
\beta\gamma}\right)+
f^{ace}f^{bde}\left(g^{\alpha\beta}g^{\gamma\delta}-g^{\alpha\delta}g^{
\beta\gamma}\right)+
f^{ade}f^{bce}\left(g^{\alpha\beta}g^{\gamma\delta}-g^{\alpha\gamma}g^{
\beta\delta}\right)
  \Big]
\\[5mm]
  +4i \left[\int \frac{dz}{z}e^{-\Phi(z)}\frac{dz'}{z'}e^{-\Phi(z')}
    V_1(z)V_2(z)G_{\left(k_1+k_2\right)^2}\left(z,z'\right)V_3(z')V_4(z')\right]
   f^{abe}f^{ecd}\cdot
\\[2mm] 
\cdot\left[g^{\alpha\gamma}k_1^{\beta}k_3^{\delta}-g^{\beta\gamma}k_2^{\alpha}
k_3^{\delta}+g^{\beta\delta}k_2^{\alpha}k_4^{\gamma}-g^{\alpha\delta}k_1^{\beta}
k_4^{\gamma}
+g^{\alpha\beta}\left(k_1^{\gamma}k_2^{\delta}-k_1^{\delta}k_2^{\gamma}
\right)+g^{\gamma\delta}\left(k_3^{\alpha}k_4^{\beta}-k_3^{\beta}k_4^{\alpha}
\right)
+\frac{1}{4}g^{\alpha\beta}g^{\gamma\delta}(k_1-k_2)^\bot (k_3-k_4)^\bot \right]
\\[5mm]
+i  \left[\int \frac{dz}{z}e^{-\Phi(z)}\frac{dz'}{z'}e^{-\Phi(z')}
    V_1(z)V_2(z)    G_0\left(z,z'\right)  V_3(z')V_4(z')\right] f^{abe}f^{ecd}
 g^{\alpha\beta}g^{\gamma\delta} 
\frac{\left(k_1^2-k_2^2\right)\left(k_3^2-k_4^2\right)}{\left(k_1+k_2\right)^2}
\\[5mm]
\left.
  +\Big[\mbox{two transpositions}\Big]
\right\}.
 \end{multline}}

Here $f^{abc}$ -- structure constants of ${\rm SU}(N_f)$ and
$V_i(z)\equiv V(k_i^2,z)$ are the ``bulk-to-boundary propagators'' which satisfy
(\ref{eq4}) with zero right hand side and subject to the boundary conditions
$V(k^2,\epsilon)=1$ (and appropriate zero conditions at $z=\zi$ or
$z=\infty$).

The transverse part of the last term in the fourth line is taken with respect to
$k_1+k_2=-k_3-k_4$, i.e.
\begin{equation}\label{eq_bot}
\left(k_1-k_2\right)^\bot \left(k_3-k_4\right)^\bot = \left(k_1-k_2\right)
\left(k_3-k_4\right)+
\frac{\left(k_1^2-k_2^2\right)\left(k_3^2-k_4^2\right)}{\left(k_1+k_2\right)^2}.
\end{equation}

Let us describe the meaning of these terms. The first term is obtained just
from the 4-vertex interaction of gauge fields (the graph \textbf{a} on FIG.
\ref{feyn2}), the second is the interaction via the virtual transverse
vector particle and the third one is obtained from the interaction with
longitudinal vector particle.
The three terms of  (\ref{eq5}) are actually obtained from the diagrams
\textbf{a} and \textbf{b} on FIG.~\ref{feyn2}, and the answer needs to be
symmetrized with respect to the permutations of $V$ in the second and
third terms. That is why we sum over two transpositions in
(\ref{eq5}) (diagrams \textbf{c} and \textbf{d} on FIG.~\ref{feyn2}).

The integral in the square brackets in the first term of (\ref{eq5}) has a UV
divergency when $\epsilon \rightarrow 0$. Indeed, at small $z$ $\Phi(z)\approx
0$, $V_i(z)\approx 1$ and we have $\int_\epsilon\frac{dz}{z}\sim \ln \epsilon$.
This UV divergency can be eliminated by adding a boundary conterterm to
the action \cite{Bianchi:2001kw}. In the particular case this means that
replacing $\epsilon\rightarrow\mu$, where $\mu$ is the UV regulator, is
sufficient. The second and the third terms in (\ref{eq5}) are finite, and after
renormalization the whole answer is UV finite.

\subsection{Two conserved and two longitudinal currents}

In the same way one can obtain the expression for the four-point vector
current correlator in the case of two conserved and two 
longitudinal currents.
This means that we consider the third, nonphysical polarization
of $v_\mu^a(k)$ (the third solution of $k_\mu v_\mu^a(k)=0$) and take the
limit $k^2\rightarrow 0$. Thus, the expression written below can be obtained
from (\ref{eq5}) by taking $v_\mu^a(k_3)$ and $v_\mu^a(k_4)$ as the third
polarizations and taking the limit $k_{3,4}^2\rightarrow 0$.

We denote conserved currents $v_\mu^a(k)$ like in the previous subsection,
$V^a_\mu(k,\epsilon)=v_\mu^a(k)$. For longitudinal currents
$V^a_\mu(k,\epsilon)=k_\mu v^a(k)$. The corresponding five-dimensional
holographic action can be written as follows
\begin{equation} 
S_5(\bot,\bot,\parallel,\parallel)=\frac{1}{4}w_{\alpha\beta}^{abcd}(\bot,\bot,
\parallel,\parallel)\,\, v_{1\alpha}^a v_{2\beta}^b v_{3}^c v_{4}^d.
\end{equation}
As in the previous subsection, $w_{\alpha\beta}^{abcd}$ represents the
holographic dual of four-point vector correlator in QCD,
{\small
\begin{multline}\label{eq_par}
  w_{\alpha\beta}^{abcd}(\bot,\bot,\parallel,\parallel)=\frac{1}{g_5^2}\left\{
  -\left[\int\frac{dz}{z}e^{-\Phi(z)}
    V_1(z)V_2(z)\right]
     k_3^\alpha k_4^\beta \cdot
\right.
\\[3mm]
  \cdot\Big[   
f^{abe}f^{cde}\left(g^{\alpha\gamma}g^{\beta\delta}-g^{\alpha\delta}g^{
\beta\gamma}\right)+   
f^{ace}f^{bde}\left(g^{\alpha\beta}g^{\gamma\delta}-g^{\alpha\delta}g^{
\beta\gamma}\right)+   
f^{ade}f^{bce}\left(g^{\alpha\beta}g^{\gamma\delta}-g^{\alpha\gamma}g^{
\beta\delta}\right)
  \Big]
\\[5mm]
  +2i
    \left[\int \frac{dz}{z}e^{-\Phi(z)}\frac{dz'}{z'}e^{-\Phi(z')}
V_1(z)V_2(z)\,G_{\left(k_1+k_2\right)^2}\left(z,z'\right)\right]
f^{abe}f^{ecd}\cdot
 \left(k_1+k_2\right)^2
    \left[k_3^\alpha k_4^\beta-k_3^\beta k_4^\alpha+\frac{1}{4}g^{\alpha\beta}
      \left(k_1-k_2\right)^\bot\left(k_3-k_4\right)^\bot\right]
\\[5mm]
  +i   
\left[\int\frac{dz}{z}e^{-\Phi(z)}\frac{dz'}{z'}e^{-\Phi(z')}V_1(z)G_{
\left(k_1+k_3\right)^2}\left(z,z'\right)V_2(z')\right] f^{ace}f^{edb}
\cdot\left(g^{\alpha\beta}-\frac{k_3^\alpha
k_4^\beta}{\left(k_1+k_3\right)^2}\right)
    \Big[\left(k_1+k_3\right)^2-k_1^2\Big]
\Big[\left(k_1+k_3\right)^2-k_2^2\Big]
\\[5mm]
    +i   
\left[\int\frac{dz}{z}e^{-\Phi(z)}\frac{dz'}{z'}e^{-\Phi(z')}V_1(z)G_0\left(z,
z'\right)V_2(z')\right]
  f^{ace}f^{edb}\cdot
  k_3^\alpha k_4^\beta \frac{k_1^2 k_2^2}{\left(k_1+k_3\right)^2}
\\[5mm]
\left.
  +\Big[\mbox{one transposition $3\leftrightarrow 4$}\Big]
\right\}
 \end{multline}}

One should remember that ``bulk-to-boundary'' propagator for longitudinal
particle is equal to unity: $V_\parallel(k^2,z)=1$. This is the reason why only
two $V$ are present in $z$ integrals in (\ref{eq_par}).
Thus, the first term is obtained from the graph \textbf{a} on FIG.~\ref{feyn2},
the second term represents the graph \textbf{b}. The third and fourth terms are
obtained from the graph \textbf{c}. The transposition of $k_3$ and $k_4$
gives the graph \textbf{d}. Note that longitudinal intermediate particle also
contribute to the final expression.

Thereby, formulas (\ref{eq5}) and (\ref{eq_par}) represent the holographic QCD
predictions for the four-point  vector currents correlators.
Apart from color and Lorentz structure they contain nontrivial external
momenta dependencies represented as integrals over the fifth coordinate in the
AdS space. It is instructive to analyze these dependencies in some particular
cases. For both formulas (\ref{eq5}) and (\ref{eq_par}) we consider two cases:
all external photons are real, two of them are virtual.

\section{Particular limits}

The $z$-dependent integrals in (\ref{eq5}) and (\ref{eq_par}) represent
nontrivial, non-perturbative contributions to the correlators. Apparently, they
are expressed as one-dimensional tree Feynman diagrams. There are two
types of such integrals: contact integrals (the first lines in (\ref{eq5}) and
(\ref{eq_par})) and terms representing one-particle exchange. Contact terms
diverge and should be regularized \cite{Bianchi:2001kw}. One-particle exchange
terms are finite. It is possible to calculate some particular limits of
one-particle exchange terms in hard-wall and soft-wall models explicitly.
Namely, the processes involving two real (or nearly real) external photons,
which have unity ``bulk-to-boundary'' propagators $V=1$.

\subsection{All four photons are real}

The most simple limit is to put all the external momenta on shell: $k_i^2=0$.
Note, in this case the terms in Eqs. (\ref{eq5}) and (\ref{eq_par}), which represent the
interaction of particles via the intermediate longitudinal particle vanish.
The next simplification comes from the fact that all $V_i(z)=1$. This means that
there is only one type of nontrivial $z$-integral:
\begin{equation}\label{eq_intms}
  I_s=-i\int \frac{dz}{z}e^{-\Phi(z)}\frac{dz'}{z'}e^{-\Phi(z')}
G_s\left(z,z'\right).
\end{equation}
This integral can be calculated precisely in both hard-wall and soft-wall
models. To do so, let us introduce the function
\begin{equation}\label{eq12x}
  K_s(z)=-i\int \frac{dz'}{z'}e^{-\Phi(z')}G_s\left(z,z'\right),
\end{equation}
which satisfy the equation
\begin{equation}\label{eq_K}
\left(z e^{\Phi(z)}\partial_z \frac{e^{-\Phi(z)}}{z}\partial_z+s\right)
K_s(z)=1.
\end{equation}
For the soft-wall after changing the variables $t=\Lambda^2 z^2$ and shifting
$K_s$ Eq.~(\ref{eq_K}) turns into Kummer's equation with the parameters
$a=-\frac{s}{4\Lambda^2}$, $b=0$. For the hard-wall model the solution of Eq.
(\ref{eq_K}) is just Bessel functions. For any model we write
\begin{equation}
K_s(z)=\frac{1}{s}\left\{1-V(s,z)\right\},
\end{equation}
where $V(s,z)$ coincide with $V(k^2,z)$ defined in (\ref{eq5}) and
(\ref{eq_par}). For the hard-wall and soft-wall models respectively we obtain
\begin{equation}
\begin{array}{cc}
V^{\rm hw}(s,z)=
  \frac{z}{\epsilon}\frac{J_1\left(\sqrt{s} z\right) Y_0\left(\sqrt{s} \zi
\right)- J_0 \left(\sqrt{s} \zi \right) Y_1\left(\sqrt{s} z\right)}{  J_1
\left(\sqrt{s} \epsilon \right) Y_0\left(\sqrt{s} \zi \right)-  J_0
\left(\sqrt{s} \zi \right) Y_1\left(\sqrt{s} \epsilon \right)},
&\,\,\,\,\,\,
  V^{\rm sw}(s,z)=
\frac{U\left(-\frac{s}{4\Lambda^2},0,\Lambda^2z^2\right)}{U\left(-\frac{s}{
4\Lambda^2},0,\Lambda^2\epsilon^2\right)},
\end{array}
\end{equation}
where $J_n(z)$, $Y_n(z)$ -- Bessel functions, $U(a,b,z)$ -- the confluent
hypergeometric function $U$.

Performing integration in (\ref{eq_intms}) gives for the hard-wall model
\begin{equation}
I_s^{\rm hw}=\int \frac{dz}{z}e^{-\Phi(z)} K_s(z)=
  \frac{1}{2s}\left\{\ln\frac{s \zi^2}{4}+2 \gamma
-\frac{\pi  Y_0\left(\sqrt{s} \zi\right)}{J_0\left(\sqrt{s} \zi\right)}\right\}
\end{equation}
($\gamma$ is Euler's constant). At large s $I_s^{\rm hw}$ behaves as
\begin{equation}
I_s^{\rm hw}\approx
  \frac{1}{2s}\left\{\ln\frac{s \zi^2}{4}+2 \gamma-\pi +\frac{2 \pi
}{1+\tan\left(\sqrt{s} \zi\right)} 
\right\}.
\end{equation}
The structure of $I_s^{\rm hw}$ at large $s$ is particularly remarkable. Apart from
the logarithm and the constant term in brackets it contains all the ``mass'' poles
$s=m^2\sim n^2$.

The analogous expression can be written for the soft-wall model,
\begin{equation}
I_s^{\rm sw}=\int \frac{dz}{z}e^{-\Phi(z)} K_s(z)= \frac{H_{-s/4\Lambda^2}}{2s},
\end{equation}
where harmonic number function for integer $n$ is defined as
$H_n=\sum_{k=1}^n\frac{1}{k}$. As $s\rightarrow\infty$
\begin{equation}
  I_s^{\rm sw}\approx \frac{1}{2s}
 \left(\ln \frac{s}{4\Lambda^2}+\gamma+\pi \cot \frac{\pi s}{4\Lambda^2}\right),
\end{equation}
and soft-wall model correctly reproduces Regge poles $s=m^2\sim n$. As we see,
the pole structure is different in each model, but large $s$ dependence remains the
same. This means that one can calculate large $s$ dependence in ``any-wall''
model (with any $z$ dependence in IR region) and it will coincide  with
hard/soft-wall model results, but the pole structure may be different.

It could be also useful to calculate (\ref{eq_intms}) for slightly virtual
particles, but computation gives the same $\ln s$ dependence, where $s$ can be
any $k^2$.

\subsection{Only two photons are real}

In the limit $k_1^2=x$, $k_2^2=y$, $k_3^2=k_4^2=0$ it is not possible to
compute all the $z$ integrals in (\ref{eq5}) and (\ref{eq_par}) even in the
large $s$ limit. But one can take the integral of type
\begin{equation}\label{eq_VGV}
  I_s(x,y)=
-i\int \frac{dz}{z}e^{-\Phi(z)}\frac{dz'}{z'}e^{-\Phi(z')} V(x,z)
G_s\left(z,z'\right) V(y,z').
\end{equation}
It can be computed in both models. To do so, let us introduce the function
\begin{equation}\label{eq_Kv}
  K_{s,y}^{(V)}(z)=-i\int \frac{dz'}{z'}e^{-\Phi(z')} G_s\left(z,z'\right)
V(y,z'),
\end{equation}
which satisfy the Eq.~(\ref{eq_K}) with $V(y,z)$ at right hand side instead
of 1. The solution for the equation
\begin{equation}\label{eq_Kkk}
  K_{s,y}^{(V)}(z)=\frac{s}{s-y} K_s(z)+\frac{y}{y-s} K_{y}(z)
\end{equation}
is a linear combination of $K(z)$ defined in (\ref{eq12x}). Now, the integral
(\ref{eq_VGV}) can be computed:
\begin{multline}
  I_s(x,y)=
\int \frac{dz}{z} e^{-\Phi(z)} V(x,z) K_{s,y}^{(V)}(z)
\\[3mm]
=-i\frac{s}{s-y} \int \frac{dz}{z} e^{-\Phi(z)}\frac{dz'}{z'} e^{-\Phi(z')}
 V(x,z) G_s(z,z')
-i\frac{y}{y-s} \int \frac{dz}{z} e^{-\Phi(z)}\frac{dz'}{z'} e^{-\Phi(z')}
V(x,z) G_{y}(z,z').
\end{multline}
Repeating again the trick in Eqs. (\ref{eq_Kv}), (\ref{eq_Kkk}) we obtain
\begin{equation}\label{eq_sqp}
  I_s(x,y)=
  \frac{s^2}{\left(s-x\right)\left(s-y\right)} I_s+
  \frac{x^2}{\left(x-s\right)\left(x-y\right)} I_x+
  \frac{y^2}{\left(y-s\right)\left(y-x\right)} I_y.
\end{equation}

It is worth mentioning that the last expression is symmetric in $s$, $x$ and $y$.

\section{Conclusions}

In the present paper we made the predictions for the strong coupling limit of four-point
vector correlators in QCD from AdS/QCD. The final expressions are represented by
formulas (\ref{eq5}) and (\ref{eq_par}). The answer includes the nontrivial external momenta dependent factors, which are
represented as integrals over the fifth coordinate. We have also computed some
particular on-shell and large-energy limits and found that these integral
factors correctly reproduce meson spectrum at small and at large energies.

The nonzero expression (\ref{eq_par}) suggests naively a nonzero possibility of
emitting real longitudinal photons. This is not the case
\cite{Gorsky:1989qd,Smilga:1990uq}. The 5-dimensional gauge invariance prohibits
this. The explanation of the paradox is related to the fact that the limit of
longitudinal photon zero virtuality is not smooth. The virtual longitudinal photon
processes contribute to the full amplitude and cross-section. Such processes in
principle could be detected in the experiment of deep inelastic scattering but
the smallest observable photon virtuality is of order $E/L$ with $L$
being the typical apparatus size.

The $\log s$ dependence does not coincide with what was obtained in
\cite{McGreevy:2007kt} for planar ${\cal N}=2$ super Yang-Mills theory.

It will be interesting to make the similar derivations for the axial four-point
correlators as it requires including the contribution of the
axial current interaction with the scalar field. Moreover, the Chern-Simons term may also
give contribution. See also recent papers \cite{Yoshida:2009dw,Bartels:2004mu,Kirschner:1983di,Hambye:2005up,Hambye:2006av} on the subject.

\section*{Acknowledgments}
I am grateful to Alexander Gorsky for stimulating discussions and supervision in
this work. I am also grateful to Andrei V. Smilga for useful discussions. I
would like to thank Olga Driga for careful reading the manuscript.
This work was supported in part by grants PICS-07-0292165, RFBR-07-02-01161 and
grant of leading
scientific schools NSH-3036.2008.2.

\end{document}